\documentclass{article}
\usepackage{amssymb,graphicx,epsfig}

\usepackage{epsfig}

\newcommand{\Bf}{\mathbf}
\newcommand{\nn}{\nonumber}
\newcommand{\eps}{\varepsilon}

\newtheorem{proposition}{Proposition}
\newtheorem{lemma}{Lemma}
\newtheorem{corollary}{Corollary}
\newtheorem{definition}{Definition}
\newtheorem{example}{Example}
\newtheorem{proof}{Proof}

\begin{document}
\title{Strongly hyperbolic systems in General Relativity~\footnote{Work supported
             by CONICOR, CONICET and Se.CyT, UNC}}

\author{
   {\sc Oscar A. Reula}
  \thanks{Member of CONICET.}\\
  {\small FaMAF, Medina Allende y Haya de la Torre,}\\
  {\small Ciudad Universitaria, 5000 C\'ordoba, Argentina}\\
}

\maketitle 
\vspace{-.4in}

\begin{abstract}

We discuss several topics related to the notion of strong hyperbolicity which are of 
interest in general relativity. 
After introducing the concept and showing its relevance we provide some covariant 
definitions of strong hyperbolicity. 
We then prove that is a system is strongly hyperbolic with respect to a given hypersurface,
then it is also strongly hyperbolic with respect to any near by one.

We then study for how much these hypersurfaces can be deformed and discuss then causality,
namely what is the maximal propagation speed in any given direction.
In contrast with the symmetric hyperbolic case, for which the proof of causality  
is geometrical and direct, relaying in energy estimates, 
the proof for general strongly hyperbolic systems is indirect 
 for it is based in Holmgren's theorem.

To show that the concept is needed in the area of general relativity we discuss
two results for which the theory of symmetric hyperbolic systems shows to be insufficient. 
The first deals with the hyperbolicity analysis of systems which are second order in space 
derivatives, they include certain versions of the ADM and the BSSN families of equations. 
This analysis is considerably simplified by 
introducing pseudo-differential first order evolution equations. 
Well posedness for some members of the latter family systems is established by showing 
they satisfy the strong hyperbolicity property. 
Furthermore it is shown that many other systems of such families are
only weakly hyperbolic, implying they should not be used for numerical modeling.

The second result deals with systems having constraints.
The question posed is which hyperbolicity properties, if any, are inherited
from the original evolution system by the subsidiary system satisfied by the 
constraint quantities. The answer is that, subject to some condition on the
constraints, if the evolution system is strongly hyperbolic then the subsidiary
system is also strongly hyperbolic and the causality properties of both are
identical.

\end{abstract}

\maketitle 

\section{Introduction}

Consider a linear first order evolution system of differential equations in $R^{n+1}$ 
on a set of tensor fields $\tilde{\phi}^{\alpha}$,
$$
\label{eq:textbook}
\partial_t \tilde{\phi}^{\alpha} = A^{i\alpha}{}_{\beta}\partial_i \tilde{\phi}^{\beta} + B^{\alpha}{}_{\beta}\tilde{\phi}^{\beta},
$$
where Greek indices denote a set of abstract indices of tensorial nature, we denote the 
dimension of that space by $m$, and
$A^{i\alpha}{}_{\beta}$, and $B^{\alpha}{}_{\beta}$ are fields on $R^{n+1}$.

The textbook definition of strongly hyperbolic systems is 
(\cite{Taylor81},\cite{Kreiss89},and \cite{Taylor91}):

\begin{definition} 
\label{def:textbook}
The above system is {\bf strongly hyperbolic} if for any co-vector $\omega_i$ in $R^n$, the matrix
$A^{i\alpha}{}_{\beta} \omega_i$ has purely real eigenvalues and is diagonalizable.
\end{definition}

For constant coefficient systems this is a necessary and sufficient condition 
for the system to be well possed in the $L^2$ sense.
Indeed, assuming constant coefficients and 
taking the Fourier transform in $R^n$ of the fields $\tilde{\phi}^{\alpha}$, which
we shall call $\hat{\phi}^{\alpha}$ 
the solutions of the resulting system of linear ordinary differential equations are:

$$
\hat{\phi}^{\alpha}(t,\omega) = (e^{(i A^i \omega_i + B)t})^{\alpha}{}_{\beta}\hat{\phi}^{\beta}(0,\omega)
$$

and so, from Plancherel identity it follows that the solution will be bounded in $L^2$ with a
bound that depends only on the $L^2$ norm of the initial data if and only if 

$$
(e^{(i A^i \omega_i +B)t})^{\alpha}{}_{\beta}
$$
is bounded independently of $\omega$. This is the case \textbf{if and only
if} the matrices $A^i \omega_i$ have purely real eigenvalues and are diagonalizable.

A matrix $A^{\alpha}{}_{\beta}$ has all its eigenvalues real and is diagonalizable 
\textbf{if and only if} there exists a positive bilinear form
$H_{\alpha \beta}$ such that $H_{\alpha \gamma}A^{\gamma}{}_{\beta}$ is symmetric.
In our case, we then have that for each given $\omega_i$ there would exists such 
$H(\omega_i)_{\alpha \beta}$ 
which would symmetrize $A^{i \alpha}{}_{\beta} \omega_i$, namely 
$H(\omega_i)_{[\alpha |\beta|}A^{i \beta}{}_{\gamma]} \omega_i=0$.

If one can find a common symmetrizer, 
$H_{\alpha \beta}$, for all $A^{i\alpha}{}_{\beta}\omega_i$ then we say 
that the system is \textbf{symmetric hyperbolic}. 

In order to understand the difference between these two concepts, 
note that any linear combination of symmetric bilinear forms remains symmetric, 
but a linear combination of diagonalizable (or symmetrizable) matrices does not necessarily remains 
diagonalizable (or symmetrizable), for instance, $\Bf{A} + \lambda\Bf{B}$, where

\[
  \Bf{A} = \left( 
    \begin{array}{cc}
         1 & 1  \\
         0 & 2  
    \end{array}
    \right)
    \;\;\;\;\; \;\;\;\;\; \;\;\;\;\;
  \Bf{B} =  \left( 
    \begin{array}{cc}
          1 & 0 \\
          0 & 2 
    \end{array}
    \right)
\]
is diagonalizable for all
$\lambda \neq -1$, but not for $\lambda=-1$.
This implies that for strong hyperbolicity it is not enough to check that the matrices corresponding 
to some basis of $\omega_i$'s are diagonalizable, one
has to check this for every co-vector $\omega_i$.

Most systems in physics are symmetric hyperbolic.
Actually it is not so easy to step into a genuinely strongly hyperbolic
system, that is one which is not also symmetric hyperbolic. In particular, if
$n=1$ all systems are symmetric hyperbolic.
So here is an example:

\begin{example}
\label{eg:Maxwell}
\begin{eqnarray}
\label{eqn:Maxwell}
\partial_t E_i &=&  \partial^j W_{ji} - (1+\alpha) \partial^j W_{ij} + \alpha \partial_i W_s{}^s \nonumber \\
\partial_t W_{ij} &=& \partial_i E_j + \beta/2 e_{ij} \partial^j E_j,
\end{eqnarray}
where Latin indices reffer to tensors on $R^3$, 
$e_{ij}$ denotes the Euclidean metric there. Here, $\phi^{\alpha} = (E_i,W_{ij})$. 
This system is strongly hyperbolic for all values of the paremeters $\alpha$ and $\beta$ such that
$\alpha \beta > 0$, but there is no symmetrizer independent of $\omega_i$ for 
$\alpha > 0$ or $\beta > \frac{-2}{3}$.
This system, subject to some constraints and gauge condition, is equivalent to Maxwell's equations, 
in vector potential form with $W_{ij}:= \partial_i A_j$. 
This is the analog of what it is done in general relativity to obtain first order systems
out of Einstein's equations.~\footnote{There are other approaches where also the Bianchi identities
are used to produce such systems, but now the variables include the Weyl Tensor.}

\end{example}


The above definition seems to depend on a particular choice of time coordinate or foliation 
on the space-time.
If that were the case this would be a way of choosing a preferred time function among 
all possible, and we know, on physical grounds, this can not be the case. 
So we must find out a covariant definition and explore
the freedom left in choosing time functions. This we do in the next section. 

The above definition of strong hyperbolicity can be extended to general quasi-linear systems, 
where now the fields $A^{i\alpha}{}_{\beta}$, and $B^{\alpha}{}_{\beta}$ depend also on the unknowns,
$\phi^{\alpha}$.
In this more general setting  strong hyperbolicity is necessary for well posedness, 
but for sufficiency one also needs smoothness of the matrix which diagonalizes the system,
$H(\phi^{\gamma},\omega_i)_{\alpha \beta}$ 
which in general depends on $\omega_i$, the space-time point and the unknowns. 
We prefer here not to include, in the definition of
strong hyperbolicity, this smoothness condition for we do not know whether it is truly 
needed, and so fundamental or it is just a temporary technical requirement for the proof
--in fact we now we do not need it in the case of constant coefficients.

How do we extend this concept to the case of manifolds? If we assume the manifold to be para-compact
then we can assume we can work on every coordinate patch, extend the functions and equations on
them to $R^{n+1}$ and stablish well posedness on each one of them using pseudo-differential
calculus. 
If in doing so we can get uniqueness and finite propagation speeds  
then we can patch the solutions found on each chart to get a global solution.
The use of Fourier transforms seems to indicate one is dealing with very specific properties
which might be valid only for particular manifolds (where we can apply Fourier series or transforms).
This is not the case, what one actually is looking is at the high frequencies limit of the solutions
of the given evolution equation and checking that in that limit they do not grow unbounded with a
growth rate which increases with the frequency. 
The behavior of bounded frequency solutions is not relevant for well posedness.
This intuitive reasoning is backed by actual theorems arising from the pseudo-differential calculus
technique, but in this article we shall try to stay away from them  and only mention the relevant 
results.
In the third section we shall look at the above problem, 
namely that of the causality properties of strongly hyperbolic
systems, to check whether it is possible or not to use this definition in non-trivial manifolds. 

The last two sections shall present applications of this theory to two problems in general relativity.
In the first we show how to analyze the hyperbolicity properties of systems which are second order in 
space derivatives, they include certain versions of the ADM and the BSSN families of equations.
This analysis is considerably simplified by 
introducing pseudo-differential first order evolution equations.
We show how to obtain first order pseudo-differential equations for these systems and then   
how to check for strong hyperbolicity on them. 
These results have already been presented elsewhere and here are only 
sketched as illustrations of the theory.
Well posedness for some members of the latter family systems is established by showing 
they satisfy the strong hyperbolicity property. 
Furthermore it is shown that many other systems of such families are
only weakly hyperbolic, implying they should not be used for numerical modeling.

The second is the study of constrained systems, on those systems, besides the evolution equations,
there are extra equations which constitute differential relations that the initial data must satisfy 
in order to produce a valid solution. 
For that to happens some integrability conditions must be satisfied, in particular
some providing the time derivatives of the quantities expressing the violation  of 
these constraints in terms of themselves and their first derivatives, so that if they 
are zero at a given time they remain so for a finite time. 
Thus, to stablish constraint preservation, an understanding of the hyperbolicity 
and causality properties of the evolution equations of these constraint quantities 
 is needed. This is also of relevance for understanding
the initial-boundary-value problem in the presence of constraints.
In the last section we shall see
which properties are inherited on the subsidiary system from the original evolution system.
It turns out that the solution here is in terms of strongly hyperbolic systems:
Strongly hyperbolic systems have strongly hyperbolic subsidiary systems.
Furthermore some causality properties also are inherited.
There are examples of systems whose evolution is symmetric hyperbolic
but whose subsidiary system is only strongly hyperbolic, showing that also for this
issue the concept of strongly hyperbolicity is needed.

\section{Strong Hyperbolicity}

We want now to introduce new definitions that reduce to the above one when a foliation
is chosen, but which do not depend on any foliation, but rather in a property of an
open set of vectors. We shall prove latter that these definitions are equivalent.


 Let 

$$
\label{eqn:evolution}
A^{a\alpha}{}_{\beta} (\nabla \phi)_a^{\beta} = J^{\alpha}
$$

be a first order system on a fiber bundle whose fiber has tangent space of dimension $m$
and whose base manifold is a space-time, $M$, of dimension $n+1$.~\footnote{
Here we are using the notation of references \cite{rG96}.} 

\begin{definition}
\label{def:A}

We shall say that this system is \textbf{A-strongly-hyperbolic} if there exists
$n_a \in T_{\star}M$ such that:

\textsl{i)} $ A^{a\alpha}{}_{\beta} n_a$
is invertible, and 

\textsl{ii)}for each (non-contractible loop on the corresponding projective space) 
$\kappa(\lambda)_a = \lambda n_a + \omega_a$ 
where $\omega_a$ is any other vector not proportional to $n_a$, \\
$dim(span\{\cup_{\lambda \in R} Kern\{ A^{a\alpha}{}_{\beta} \kappa_a(\lambda)\}\}) = m$.

\end{definition}

\noindent \textbf{Remarks:}

\begin{enumerate}

\item Choose a vector $t^a$ such that $t^a n_a =1 $ and define 
$h_a{}^b = \delta_a{}^b - n_at^b$ then the above equation becomes, 

$$
A^{a\alpha}{}_{\beta}n_a t^b (\nabla \phi)_b^{\beta} 
= - A^{a\alpha}{}_{\beta} h_a{}^b(\nabla \phi)_b^{\beta} + J^{\alpha}.
$$

Since $A^{a\alpha}{}_{\beta}n_a $ is invertible we recuperate the textbook definition by
multiplying by its inverse,

$$
t^b (\nabla \phi)_b^{\alpha} 
= (A^{a\alpha}{}_{\beta}n_a)^{-1} (A^{a\beta}{}_{\gamma} h_a{}^b(\nabla \phi)_b^{\gamma} + J^{\beta}).
$$

The diagonalizability of 
 $(A^{a\alpha}{}_{\beta}n_a)^{-1} A^{b\beta}{}_{\gamma}\omega_b$
follows from the definition above, for it implies that the above matrix has a complete set 
of eigenvectors. Furthermore all its eigenvalues are real.

\item Again it is not true that if $\Bf{B}$ is a diagonalizable and invertible matrix and $\Bf{A}$ is
diagonalizable, then $\Bf{B}^{-1} \Bf{A}$ is diagonalizable. The matrices of the example above are a
counterexample. 

\item In examples (e.g. the first order version of the wave equation in more than two space-time 
dimensions) it is easy to see that not every $n_a$ for which 
$A^{a\alpha}{}_{\beta}n_a$ is invertible is a good candidate
for property \textsl{ii)}, in fact it seems that only a connected subset of them in 
$P^n$ (see item 5 below) comprises the whole set of possible $n_a$'s.

\item This definition still depends on a particular $n_a$ and we must prove that the set
of allowed $n_a$'s is open. Otherwise it is a very restrictive definition, not better
than definition \ref{def:textbook}.

\item We should think on the loops above on the projective space $P^n$, where any two vectors are
identified if they are parallel. This is a compact non-simply connected space. 
The $\kappa_a(\lambda)$ loops are not contractible to zero, are compact, and continuous with respect
to $\lambda, n_a$ and $\omega_a$. This implies, 

\begin{itemize}

\item The roots of the determinant of $A^a \kappa_a$ are continuous functions of the arguments.
So the set of points in $P^n$ where the kernel of $(\lambda n_a + \omega_a)A^a$
is non-trivial is a continuous manifold of dimension $n-1$. Furthermore, since 
$(A^{a\alpha}{}_{\beta}n_a)^{-1} A^{b\beta}{}_{\gamma}\omega_b$ is diagonalizable for any
$\omega_b$, and perturbations of eigenvalues due to diagonalizable matrices are Lipschitz,
such manifold is Lipschitz. 
This sub-manifold can have some self-intersections where two or more eigenvalues coincide.

\item  The curves $\kappa(\lambda)_a:= \lambda n_a + \omega_a$ pinches off transversally this manifold.
This follows easily from the fact that if we move continuously $\omega_a$ the roots $\lambda_i$
must also move  continuously and can not disappear, (become complex), because they are a complete
set and are all real, so they must just move along the real axis. This implies that if we
slightly perturb $n_a$ the new curve must also transverse all these manifolds without 
loosing any root. The problem is then to assert that one does not looses any subspace in
the kernel. 

\item If we move continuously $n_a$ the (real) zeroes can not dissappear (become complex) so 
at most can move  towards infinity, but those are precisely the points where 
$n_a$ is no longer invertible.

\item $f(n_a,\omega_a) := dim(\cup_{\lambda \in R} Kern\{ K^{a\alpha}{}_{\beta} \kappa_a(\lambda)\})$
is a upper-semi-continuous function of its arguments. That is, given any point $(n_a,\omega_b)$ there
is a always a neighborhood $V$ such that 
$f(\tilde{n}_a,\tilde{\omega}_a) \leq f(n_a,\omega_b) \;\;\forall\; (\tilde{n}_a,\tilde{\omega}_a) \in V$
\end{itemize}

\end{enumerate}

We are now ready to prove the main result of this section:

\begin{proposition}
\label{prop:1}
If for some $n_a$ properties (\textit{i, ii}) hold, then they also hold
for any other co-vector in a sufficiently small open neighborhood of $n_a$.

\end{proposition}

\begin{proof} It is useful to think the problem (and use the corresponding topology)
in real projective space $P^n$, where two vectors are identified if they are parallel. 
In that case the image of the curves $\lambda n_a + \omega_a$ are compact 
non-contractible curves and so we can define a topology between them in a trivial way.

We assume properties (\textit{i, ii}) hold for some given $n_a$, and --for contradiction-- that
they do not hold for some  $\tilde{n}_a = n_a + \eps \delta_a$, $\eps > 0$ with $\eps$ 
sufficiently small. That means that at some root $\tilde{\lambda}_i$, the matrix
$(\tilde{\lambda}_i \tilde{n}_a + \omega_a)A^a$, has a kernel with lower dimensionality
than the one at $(\lambda_i n_a + \omega_a)A^a$ where $\lambda_i$ is the corresponding
root for this curve. 
Consider now the curve given by 
$\lambda n_a + (\omega_a + \tilde{\lambda_i} \eps \delta_a)$. 
At $\lambda = \tilde{\lambda}_i$ this curve intersects the above one and so its kernel at
this point has also lost dimensions compared with the point $(\lambda_i n_a + \omega_a)$
Since properties (\textit{i, ii}) hold for this new curve there must be at least another point,
$\lambda_j n_a + (\omega_a + \tilde{\lambda_i} \eps \delta_a)$ at which the kernel has increased
its dimensionality compared with the nearby point $\lambda_j n_a + \omega_a$ where the first
curve has a non-trivial kernel. But this is impossible, since the dimension of the kernel is
upper semi-continuous.
\end{proof}


\begin{proposition}
\label{prop:2}
 Let $n_a$ a co-vector such that properties (\textit{i, ii}) hold, let 
$S(n_a)$ be the connected component of vectors satisfying (\textit{i, ii}), then the 
boundary of $S(n_a)$ consists of vectors $m_a$ where $A^a m_a$ ceases to be invertible.
\end{proposition}

\begin{proof} Consider a one parameter family of vectors 
$v_a(\lambda):=n_a + \lambda\delta_a$. Let $\lambda_0$ be the least upper
bound of
$\{\lambda | \mbox{given any co-vector } \omega_a \mbox{the curve } \lambda v_a(\lambda) + \omega_a $ $\mbox{has
m real roots (counting multiplicities)} \}$. For $\lambda \in [0,\lambda_0)$ 
the roots are real, and so by 
continuity the roots of the curve at $\lambda_0$ must be also real, and so the only 
possibility to loose
a root is that it goes to infinity, that means the $v_a(\lambda_0)A^a$ has non-trivial kernel.
We need to see now that this is a boundary point, that is, that at no point before $\lambda_0$ the 
dimension of the direct sum of the kernels is not smaller than $m$. Assume, for contradiction
this happens, namely there is $0 < \lambda_1 < \lambda_0$ being the least upper bound of the 
set of $\lambda$'s
for which the dimension of the direct sum of kernels is $m$. Since the Kernel is a upper 
semi-continuous function at the point where it lowers its value must retain its upper value.
So at $v_a(\lambda_1)$ the dimension of the direct sum is $m$, but in arbitrarily near points
that sum is smaller. But this contradicts that the set for which that sum is $m$ is open.
Unless at $\lambda_1$ the matrix ceases to be invertible.
\end{proof}


We know (see for instance \cite{Kreiss89}) that $\tilde{A}$ is a diagonalizable matrix 
with real eigenvalues if and only if the following inequality holds:

$$
|(sI + \tilde{A})^{-1}| \leq K / |Im(s)| \;\;\;\; \forall s \in C, \;\; Im(s) \neq 0 
$$

Thus, if a system is strongly hyperbolic for $n_a$, then, since $A^an_a$ is invertible
and $(A^an_a)^{-1} A^a\omega_a$ is diagonalizable with real eigenvalues. Thus we have,

\begin{eqnarray}
\label{eqn:kreiss_ineq}
|(sA^an_a + A^a\omega_a)^{-1}| &=& |(sI+ (A^an_a)^{-1}A^a\omega_a)^{-1}(A^an_a)^{-1}| \nn \\
                        &\leq & |(A^an_a)^{-1}| |(sI + (A^an_a)^{-1}A^a\omega_a)^{-1}| \nn \\
                        &\leq & |(A^an_a)^{-1}| K/|Im(s)| .
\end{eqnarray}

The reverse inequality is also true, thus, the above inequality is true if and only if
$\tilde{A}:=(A^an_a)^{-1} A^a \omega_a$ is diagonalizable and all its eigenvalues are real.
So we can use also this property as an alternative definition of strong hyperbolicity. 

\begin{definition}
 The above system is \textbf{B-strongly hyperbolic} if there exists $n_a$ such that
for any $\omega_i$ and any $s \in C$ with $\Im s \neq 0$ we have,

\begin{eqnarray}
\label{eqn:kreiss_ineq_2}
|(sA^an_a + A^a\omega_a)^{-1}|  &\leq & \tilde{K}/|Im(s)| .
\end{eqnarray}
\end{definition}

\begin{proposition}
A-strong hyperbolicity and B-strong hyperbolicity are equivalent.
\end{proposition}

Let us see that this is indeed the inequality needed for proving well posedness.
We consider the constant coefficient case and take a constant flat foliation, make Fourier 
transform in space 
$$
\tilde{\phi}(t,\omega_i) = \frac{1}{\sqrt{2\pi}^n} \int e^{-i\omega_j x^j}\phi(t,x^j) dV
$$
and then a Laplace transform in time,
$$
\hat{\phi}(s,\omega_i) = \int_0^{\infty} e^{-st}\tilde{\phi}(t,\omega_i) dt \;\;\;\;\; \Re{s} > \alpha > 0.
$$

Thus Eq. ( \ref{eq:textbook}) becomes, 

$$
A^0(s \hat{\phi}(s,\omega_l) - \tilde{\phi}(0,\omega_l)) + iA^j \omega_j \hat{\phi}(s,\omega_l) = 0
$$

Thus we have, 

$$
\hat{\phi}(s,\omega_l) = (A^0 s + i A^j \omega_j)^{-1} A^0 \tilde{\phi}(0,\omega_l)
$$

Using Plancherel identity we get, with $\eta > \alpha$,

\begin{eqnarray}
\int_0^T ||\phi(t,\cdot)||^2 dt 
          &\leq& 
                \frac{e^{2 \eta T}}{(2\pi)^{n/2}} \int^{\infty}_{\infty} 
                ||(A^0 (\eta + i \zeta) + i A^j \omega_j)^{-1} A^0 \tilde{\phi}(0,\omega_l)||^2 
                d\zeta d\omega^n \nn \\
          &\leq& 
                \frac{e^{2 \eta T}}{(2\pi)^{n/2}} \int^{\infty}_{\infty} 
                ||A^0 \tilde{\phi}(0,\omega_l)||^2 
                ||(A^0 (\eta/\omega + i \hat{\zeta}) + i A^j \hat{\omega}_j)^{-1}||^2 
                \omega^{-2} d\zeta d\omega^n \nn \\
          &\leq& 
                \frac{K^2 e^{2 \eta T}}{(2\pi)^{n/2} \eta^2} \int^{\infty}_{\infty} 
                ||A^0 \tilde{\phi}(0,\omega_l)||^2 d\zeta d\omega^n \nn \\
          &\leq&
                \frac{K^2 e^{2 \eta T}}{(2\pi)^{n/2} \eta^2}||A^0||^2 ||\phi(0,\cdot)||^2
\end{eqnarray}

where in the last inequality we have used inequality (\ref{eqn:kreiss_ineq}), namely,  \\
$||(A^0 (\eta/\omega + i \hat{\zeta}) + i A^j \hat{\omega}_j)^{-1}||^2 \leq K\omega/\eta 
\;\;\forall \;\; \hat{\omega}_i \in S^{n-1}$.


An alternative proof of proposition \ref{prop:1} is the following: assume inequality (\ref{eqn:kreiss_ineq})
holds for some $n_a$ and all $\omega_a$ and try to probe it for $\tilde{n}_a = n_a + \eps 
\delta_a$ where $\eps$ is taken small enough.

We have:

\begin{eqnarray}
A^a(s\tilde{n}_a + \omega_a) &=& A^a( s_R (n_a + \eps \delta_a) + \omega_a) + i s_I A^a (n_a + \eps \delta_a) \nn  \\
                      &=& A^a n_a(T \Gamma T^{-1} + i s_I (I + \eps (A^a n_a)^{-1}A^a\delta_a)) \nn  \\
                      &=& A^a n_a T (\Gamma + i s_I (I + \eps T^{-1}(A^a n_a)^{-1}A^a\delta_a T))T^{-1}
\end{eqnarray}
Here $T\Gamma T^{-1} = (A^a n_a)^{-1}(s_R I + A^a(\omega_a + \eps \delta_a))$ where $\Gamma$ is 
a diagonal matrix with all entries real. 
Such decomposition is possible because we have applied the assumption that 
$(A^a n_a)^{-1} A^a\tilde{\omega}_a$ is diagonalizable and has purely real eigenvalues 
for $\tilde{\omega}_a = \omega_a + \eps \delta_a$.
$\Gamma$ and $T$ depend on $s_R$, but since $T$ only depends on the direction of  
$s_R (n_a + \eps \delta_a) + \omega_a$ and the set of possible directions is a compact set, then
$|T|$ and $|T^{-1}|$ are bounded.

Thus, 
\begin{eqnarray}
(A^a(s\tilde{n}_a + \omega_a))^{-1} 
&=& T (\Gamma +   i s_I (I + \eps T^{-1}(A^a n_a)^{-1}A^a\delta_a T))^{-1}T^{-1}(A^a n_a)^{-1} \nn \\
&=& \frac{1}{is_I} T \{  \sum_{n=0}^{\infty} (is_II + \Gamma)^{-n} (is_I)^n (-1)^n  B^{n-1} \} T^{-1}(A^a n_a)^{-1}
\end{eqnarray}
where $B = \eps T^{-1}(A^a n_a)^{-1}A^a\delta_a T$
But since $\Gamma$ is diagonal and real we have that $|s_I (is_II + \Gamma)^{-1}|$ is bounded and so for
sufficiently small $\eps$ the series converges and so the norm of the resolvent is bounded by some 
constant times 
$1/|Im(s)|$.

The second proposition can also be proven using Eq. (\ref{eqn:kreiss_ineq}).


Consider the following alternative definition (due to R. Geroch \cite{rG02}):

\begin{definition} 
A system is C-strongly hyperbolic if:

\textsl{i)} For each $\omega_a$ there exists $h_{\beta \gamma}(\omega_a)$ such that 
$\tilde{A}_{\beta \alpha}:= h_{\beta \gamma}(\omega_a) A^{a\beta}{}_{\alpha} \omega_a$ 
is symmetric, and

\textsl{ii)} For some $n_a$, and any $\omega_b$, $h_{\beta \gamma}(\omega_b) A^{a\beta}{}_{\alpha} n_a$ 
is symmetric and positive definite.
\end{definition}


We want to show that this definition is equivalent to the one already introduced.

\begin{proposition}
\label{prop:4}
A-strong hyperbolicity and C-strong hyperbolicity are equivalent.
\end{proposition}

\begin{proof}
Assume the system is A-strongly hyperbolic. 
The kernel condition implies that for any $\omega_c$ 
$(A^a n_a)^{-1}{}^{\beta}{}_{\gamma} A^{c \gamma}{}_{\alpha}\omega_c$ 
has a complete set of eigenvalues,
thus there exists a positive bilinear form $H_{\delta \beta}(\omega_c)$ such that 
$H_{\delta \beta}(\omega_c) (A^a n_a)^{-1}{}^{\beta}{}_{\gamma} A^{b \gamma}{}_{\alpha}\omega_b$
is symmetric. But then, taking 
$h_{\delta \alpha}(\omega_c):= H_{\delta \beta}(\omega_c) (A^a n_a)^{-1}{}^{\beta}{}_{\alpha}$
we have \\ 
$h_{\delta \alpha}(\omega_c)(A^{b\alpha}{}_{\gamma}n_b = H_{\delta \gamma}(\omega_c)$, and
$h_{\delta \gamma}(\omega_c)A^{b\gamma }{}_{\alpha}\omega_b = H_{\delta \beta}(\omega_c) (A^a n_a)^{-1}{}^{\beta}{}_{\gamma} A^{b \gamma}{}_{\alpha}\omega_b$
and so both are symmetric and the first is positive definite.
 
Assume now we have a C-strongly hyperbolic system. 
We want to show that 
for each (non-contractible loop on the corresponding projective space) 
$\kappa(\lambda)_a = \lambda n_a + \omega_a$ 
where $\omega_a$ is any other vector not proportional to $n_a$, \\
$dim(span\{\cup_{\lambda \in R} Kern\{ A^{a\alpha}{}_{\beta} \kappa_a(\lambda)\}\}) = m$

Now, $u^{\alpha}$ is in such kernel for some $\lambda$ if and only if
$\lambda A^{a\alpha}{}_{\beta}n_a u^{\beta} = A^{a\alpha}{}_{\beta}\omega_a u^{\beta} $.
Multiplying both members by $h(\omega_c)_{\gamma \alpha}$ we get,
\[
\tilde{A}_{\gamma \delta}u^{\gamma}u^{\delta} = \lambda H_{\gamma \delta}u^{\gamma}u^{\delta}
\]
with both $\tilde{A}_{\gamma \alpha}$, and $H_{\gamma \beta}$ symmetric and the latter
positive. 
But then the eigenvalue-eigenvector pairs are obtained as minimizing vectors 
of the function 
$F(u^{\alpha}) := \tilde{A}_{\gamma \delta}u^{\gamma}u^{\delta}/H_{\gamma \delta}u^{\gamma}u^{\delta}$
on a diminishing sequence of $H$ orthogonal subspaces, from which we conclude they form a complete set of vectors. 
\end{proof}

\section{Causality}


An important property of evolution systems is the propagation speed of high frequency 
perturbations.
In particular the highest speed in any given direction, via Holmgren's 
Theorem (see for instance \cite{Taylor91})
determines the domain of dependence of the system for any given initial 
hypersurface.~\footnote{Strictly 
speaking this is only asserted for linear systems with analytic coefficients, 
but will be generalized here.} 

Given a small enough region with local coordinate system $(t,x^i)$ 
we consider fields of the form, 
$\phi^{\alpha} = u^{\alpha}_{\omega} e^{\sigma_{\omega} t - \omega_i x^ i}$, where 
$u^{\alpha}_{\omega}$ is an eigenvector (and $\sigma_{\omega}$ the corresponding eigenvalue)
of $A^{i \alpha}{}_{\beta}\omega_i$. 
Then in the limit of high frequencies these fields are very close to solutions of the 
evolution equations and have velocities given by 
$v^i = \frac{\partial \sigma}{\partial \omega_i}$. 
The surfaces tangent to $\sigma_{\omega} dt - \omega_i$ are called the characteristic surfaces
of the system and along them the respective eigen-modes propagate,
being the integral curves of $V^a = (\partial_t)^a + v^a$.
The highest of these velocities for any given $\omega_i$ (which corresponds to the 
highest eigenvalue, since the dependence of the eigenvalues with respect to
$\omega_i$ is homogeneous of degree one) determines the causality properties of the 
system.

The argument for this last assertion is as follows: 
Consider a quasi-linear strongly hyperbolic system, say (\ref{eq:evolution}), and
take a initial slice $S$ 
which can be either continued to $R^n$ or $T^n$, so that we can define on it either 
Fourier transforms or Fourier series and so introduce the pseudo-differential calculus needed to 
show well posedness for strongly hyperbolic systems. 
Extend the whole space-time to include a finite  neighborhood of that extended slice,
and  give in the extended slice initial data, $\phi_0^{\alpha}$ with enough smoothness
and decay properties as needed to use the existence theorems for 
strong hyperbolic systems (assuming the 
symmetrizer is smooth in all variables) to show that there exists a finite neighborhood 
of that initial surface where a unique solution, $\phi^{\alpha}$, exists. 
Consider now linear perturbations of that solution. 
Quasi-linearity implies they will satisfy an equation of the form,

\begin{equation}
\label{eq:perturbation}
L^{\alpha}{}_{\beta} \delta\phi^{\beta} := \partial_t \delta \phi^{\alpha} - A(\phi,t,x)^{i \alpha }{}_{\beta}\partial_i \delta \phi^{\beta} + 
B(\phi,t,x)^{\alpha}{}_{\beta}\delta \phi^{\beta}.
\end{equation}
That is, with the same principal part as the original one.

We want to show that the domain of dependence of these perturbations is characterized by the 
highest eigenvalues for each direction. 

Following the same strategy as for Holmgren's theorem \cite{Taylor91},
we take a point $p \in S$ and two global coordinate systems with origin at $p$,
$(t,x^i)$ and $(\hat t, x^i)$ which 
differ on the time coordinate in a cuadratic way near $p$ so that we have two time foliations 
with the intersection of the past of one from one fixed time on with the future of the other 
form another fixed time on form a lens shaped region  
$\Omega_{T}= \{t \geq 0\} \cap  \{\hat{t} \leq T\}$.
We shall call the past bounary $\Sigma_0$ and the future one $\hat{\Sigma}_T$.
Choose  the perturbation initial data, $\delta \phi_0^{\alpha}$ to vanish in $\Sigma_0$, 
but let it be otherwise arbitrary (but sufficiently smooth) everywhere else.
The domain of dependence of such region will then be the region of 
space-time where the perturbation remains null, 
regardless of the initial values $\delta \phi_0^{\alpha}$ takes outside 
that region. 
If $T$ is choosen sufficiently small then one can arrange for the normal to the $\hat{t}$
constant slices, $\hat{n}_a = d\hat{t}$,  to be very close to the normals to the $t$ constant slices, 
$n_a = dt$, so that with respect to both normals
the evolution systems is strongly hyperbolic.

We now consider solutions of the $L^2$ adjoint of the operator to (\ref{eq:perturbation}), 
$L^{\alpha}{}_{\beta}$
to the past of the surface $\hat{t}=T$ with vanishing initial data but with sources,

$$
L^{\star}_{\alpha}{}^{\beta}\psi^i_{\beta} = p^i{}_{\alpha}
$$
where $p^i{}_{\alpha}$ are a set of basis vectors times a function which inside the region 
$\Omega_{T}$ is a polynomial of degree $i$ in the coordinates and outside it decays exponentially
to zero in such a way as keeping these sources smooth. 
Since the adjoint operator is also strongly hyperbolic the proof of well posedness for 
these systems  will hold for each one of these source functions, so we have a dense set
of solutions $\{\psi^i_{\beta}\}$ satisfying:

$$
\int_{\Omega_{T}} p^i{}_{\beta}\delta \phi^{\beta}
= 
\int_{\Omega_{T}} L^{\star}{}_{\alpha}{}^{\beta} \psi_i{}_{\beta} \delta \phi^{\alpha}  
= 
\int_{\Omega_{T}} L^{\alpha}{}_{\beta} \delta\phi^{\beta} \psi_i{}_{\alpha} = 0
$$
where the boundary terms go away since $\phi^{\gamma}$ vanishes at the past boundary 
$\Sigma_0$, and $\psi_{\gamma}$ vanishes at the future
one, namely $\hat{\Sigma}_T$.
Thus, since the set of polynomials are dense, in $L^2$ we conclude that $\phi^{\gamma}=0$ inside the
foliation. 

The result can be extended to the full solutions of the quasi-linear system
by proving it at each step of the iteration:

\begin{equation}
\label{eqn:iteration}
 \partial_t \phi^{\alpha}_{n+1} = A(\phi_n,t,x)^{i \alpha }{}_{\beta}\partial_i \phi^{\beta}_{n+1} + 
B(\phi_n,t,x)^{\alpha}{}_{\beta}\delta \phi^{\beta}_{n+1}.
\end{equation}
Along the iteration the region where the solutions to the corresponding adjoint
equations are valid can be shown to contain a given finite time interval
which depends only on the initial data for $\phi^{\alpha}$.

Since the lens shaped regions can be extended until the normal makes $A^{a \alpha}{}_{\beta}\hat{n}_a$
singular we have the following result: 

\begin{proposition}
Given a first order quasilinear strongly hyperbolic system of equations the domain of dependence
of regions is determined by the highest characteristics speeds along each direction. 
\end{proposition}

A different argument leading to causality can be obtained for constant coefficient systems 
by directly constructing 
their solutions as limits of grid solutions to their discretized versions. 
Indeed, as argued by Kreiss~\cite{Kreiss}, after embedding the system in $R^{n+1}$ or $T^{n+1}$
one can discretize the system using the method of lines and a centered difference operator of some 
given order. 
Such operators involve a finite number of grid points.~\footnote{We refer to a \cite{Gustafsson95} for the 
basic results we are quoting.}
Thus obtaining a big system of linear ordinary differential equations. 
It is easy to check the eigenvectors form a complete set and that the eigenvalues have no positive real part.
Thus, for sufficiently small time-step size (depending on the space inter grid size) the system is stable and
a solution exists and is bounded for any finite time interval. 
But at given grid size, the value of the solution at a given position in space and time depends, by construction,
only on the values of the initial data grid functions in a finite region of the initial surface.
When increasing the resolution the number of grid points which enter in such calculation increase,
but not the size of the region of dependence, thus in the limit we see causality holds.

Thus we see that strongly hyperbolic systems share many properties with the subclass of
symmetric hyperbolic ones: 1.- One can prescribe initial data on any Cauchy surface, 
that is a surface such that its normal $n_a$ satisfies the conditions on any of the equivalent definitions of 
strong hyperbolicity we have introduced; 2.- The causality properties and propagation speeds of
high frequency perturbations are derived from the characteristic surfaces of the system, 
namely surfaces at which the normal is such that $A^{a \beta}{}_{\alpha}n_a$ is not invertible.

But, are really necessary strongly hyperbolic systems in physics? 
It seems that all classical evolution equations in physics can be described in one way or 
another by symmetric
hyperbolic systems, so why complicate matters introducing strongly hyperbolic systems?
In what follows we shall consider two examples where introducing strongly hyperbolic systems 
lead to new insights in physics.


\section{Well posedness of systems which are second order in space derivatives}


Here we sketch the analysis of well posedness for systems which are first order in 
time derivatives and second order in space derivatives. Details can be found in \cite{NOR}. 
Main examples of such systems are the ADM equations of general relativity, \cite{ADM} 
and the more recent modification of it called BSSN, \cite{SN}.

Since well posedness is a property related to the behavior of high frequency solutions
in establishing it one looks at the equations and their solutions in that limit.
It is a property of well posed systems that zero order terms do not change the properties 
of well posedness in the sense that if a system is well posed, then under the addition of a
zeroth order term the system remains well posed \cite{Kreiss89}. 
Thus, is we have a first order system we know that all we have to do is to look at
its principal part, namely the terms containing first derivatives. 
So the strategy here to analyze these second order systems is to
transform them into equivalent first order systems and then look at algebraic 
properties of their principal part.
There are several ways of obtaining a first order system out of these second order ones.
One of them, already used in \cite{SR99}, and \cite{SCPT02}, is to add as variables 
all first order derivatives and look at the resulting larger system, where second derivatives are
substituted by first derivatives of the added variables and new evolution equations are introduced
for the new variables by commuting space and time derivatives. 
This analysis has the drawback that the resulting system is considerably enlarged and 
many new constraints appear, although in \cite{SCPT02} the propagation of these new constraint
is trivial.
Another approach, the one sketched here, is to add as new variables the square root of the Laplacian
of some of the original variables and so get a first order pseudo-differential system.
The advantage of this method is that now the principal part of the pseudo-differential 
system is algebraically much simpler to deal with, and the new constraints arising 
from the introduction of the new variables are fewer and their propagation is trivial. 
This procedure has been used by Taylor to describe higher order operators, \cite{Taylor91}
in a general setting and by \cite{KO} for Einstein's equations as a fully second order system. 
We shall consider here only constant coefficient systems. To deal with constant coefficient
systems is all what it is actually needed, for if they are well possed, then well possed holds for
all systems whose ``frozen coefficient'' versions coincide with them.

We consider the ADM equations linearized off Minkowski space-time,

\begin{eqnarray}
\label{eq:ADMd1}
\partial_t \delta h_{ij} &=& -2 k_{ij} \nonumber \\
\partial_t k_{ij} 
&=& \frac{1}{2} h^{kl} \left[ - \partial_k\partial_l \delta h_{ij} 
- (1+b) \partial_i\partial_j \delta h_{kl}
+ 2 \partial_k\partial_{(i} \delta h_{j)l} \right] \nonumber
\end{eqnarray}
where before linearizing the lapse has been densitized as 
$N = (g)^b Q$ where $Q$ is considered a given function of space-time and the shift has
been set to zero.

We now introduce a new variable, $\ell_{ij} = \sqrt{\Delta}_h \delta h_{ij}$, 
and the corresponding evolution equation:

\begin{eqnarray}
\partial_t \ell_{ij} &=& -2  \sqrt{\Delta}k_{ij} 
\end{eqnarray}

Substituting in the second equation of (\ref{eq:ADMd1}) all $h_{ij}$ for the new
variable we get a first order pseudo-differential system of evolution equations.
The necessary and sufficient conditions for well posedness of this type of systems
are the same as those of differential systems. In fact the proof is the same.
Thus it is enough to look at the principal symbol, 
[Namely the limit of the right hand side of the equation where all derivatives have been substituted
by $i\omega_j$ divided by $|\omega|$ when $|\omega| \to \infty$, and subsequently multiplied back
by $|\omega|$.] which in this case is:

  \begin{eqnarray}
\tilde{A}
\left(
  \begin{array}{c}
    \hat \ell_{ij}\nonumber \\ 
    \hat k_{ij}\nonumber
  \end{array}
\right) 
&=& 
    |\omega|
    \left(\begin{array}{c}
        - 2 \hat k_{ij}  \nonumber\\
        -(1/2) \left( \hat \ell_{ij}
          +(1+b)\tilde \omega_i\tilde \omega_j h^{kl}\hat \ell_{kl}
          - 2 \tilde \omega^k \tilde \omega_{(i}\hat \ell_{j)k} \right)
       \nonumber
  \end{array}
\right) 
\end{eqnarray}
where we have substituted each partial derivative by $\omega_i$, and defined,
$\tilde \omega_i := \omega_i/|\omega|$.

Thus what we need to know is whether matrix $\tilde{A}$ can be diagonalized.
It turns out that it can not, although all eigenvalues are real.
So the system is only weakly hyperbolic.
Furthermore it can be shown that the addition of the Hamiltonian constraint does not
resolve the degeneracy (although it changes the characteristic speeds and the 
subsidiary system of equations for constraint propagation).

The BSSN equations are a further generalization of the above system in which some
linear combination of first derivatives of the metric are promoted to new independent
variables and in the corresponding new evolutions equations the momentum constraint
is conveniently added. The new variable is defined with respect to a background 
connection $\partial_i$ and is given by 
$f_i := \frac12 h^{kl}(\partial_k h_{il} - \partial_i h_{kl})$
and in its evolution equation the momentum constraint  
is added (with a new parameter $c$) as follows:

\begin{equation}
\partial_t f_i := -2 \partial_k k^k{}_i + \partial_i k + c (\partial_k k^k{}_i - \partial_i k)
\end{equation} 

In this case the principal symbol becomes:

  \begin{eqnarray}
\tilde{A}
\left(
  \begin{array}{c}
    \hat \ell_{ij}\nonumber \\ 
    \hat k_{ij}\nonumber \\
    \hat f_i \nonumber
  \end{array}
\right) 
&=& 
    |\omega|
    \left(\begin{array}{c}
        - 2 \hat k_{ij}  \nonumber\\
        -(1/2) \left( \hat \ell_{ij}
          + b\tilde \omega_i\tilde \omega_j h^{kl}\hat \ell_{kl}
          - 2 \tilde \omega_{(i} \hat f_{j)} \right)\nonumber \\
        (c-2)\tilde \omega^k \hat k_{ik} + (1-c)\tilde \omega_i h^{kj}\hat k_{kj} \nonumber
  \end{array}
\right) 
\end{eqnarray}
It turns out in this case that for non-zero values of $c$ 
the symbol is diagonalizable with all eigenvalues real, so
the system is strongly hyperbolic in the algebraic sense. Furthermore one can show that
the symmetrizer is a smooth function of all variables, so that the well posedness proof
for the fully quasi-linear, pseudo-differential case applies.
This we resume in the following \cite{NOR}:

\begin{proposition}
The BSSN family of systems with $c > 0$ is well posed.
\end{proposition}

Thus we have seen that the application of the concept of strongly hyperbolicity, 
although used here in the broader class of pseudo-differential quasi-linear systems
facilitates the well posed analysis of higher order evolution systems.


\section{Constrained Systems}


In this section we deal with another issue for which the notion of strong hyperbolicity
is critical. That is the initial value problem for constrained systems.

In physics one usually finds that the fields under consideration do not only satisfy some 
evolution 
differential equations but also are subject to some extra differential relations, called 
constraints,
 which can be thought of as relations at a given time, thus effectively diminishing the degrees of 
freedom of the theory.
In order for a theory of this type to have solutions some integrability conditions between these
two set of equations must be satisfied. 
Part of these integrability conditions basically assert that if the constraint equations are 
satisfied at some initial surface, then their time derivative off such surface must also vanish,
that is, these derivatives should be expressed as homogeneous functions of the constraints 
themselves and their derivatives tangent to the initial hypersurface. 
We shall call these evolutionary integrability conditions, \textit{the subsidiary system}; 
they are equations on the \textit{constraint quantities}, 
namely the quantities representing the constraint equations.
If the evolution equations as well as the constraint equations are first order in
derivatives, something we shall assume from now on, then these evolutionary integrability 
conditions, form also a semi-linear system of first order differential 
equations on the \textit{constraint quantities}.

The evolutionary properties of these subsidiary systems are very important.
At the continuum level in asserting whether or not the constraints will be preserved 
(continue to hold) along evolution. In particular, when boundaries are present, not all otherwise 
allowed boundary conditions are consistent with constraint evolution. In order to elucidate 
which ones can be given it is usually necessary to have a detailed knowledge of the properties 
the subsidiary system.
But even more at the discrete level, for in 
numerical simulations the constraints are never exactly satisfied and so it is necessary to
study how these errors will propagate along the simulation.
Thus, it is natural to ask which properties, if any, these subsidiary systems inherit from 
the evolution equations. This is going to be the main task of this section.
 We shall see --under certain assumption on the constraint system structure-- 
that if the evolution system is strongly hyperbolic then so is the subsidiary system,
 and furthermore the characteristics of the subsidiary system are a subset of the characteristics 
of the evolution 
system.

We next define the set of equations we are going to consider. For a more formal definition of
many of the concept here introduced see \cite{rG96}.

\begin{definition}
 By an evolution system with differential constraints we mean: 

\textit{i}) a first order evolution system of quasi-linear partial differential equations of the 
form,

\begin{equation}
\label{eq:evolution}
\partial_t \phi^{\alpha} = A(\phi,t,x)^{i \alpha }{}_{\beta}\partial_i \phi^{\beta} + B(\phi,t,x)^{\alpha}, 
\end{equation}

\textit{ii}) a first order system of quasi-linear partial differential equations
--the constraint system-- of the form,

\begin{equation}
\label{eq:constraints}
C^A := K(\phi,t,x)^{iA}{}_{\beta}\partial_i \phi^{\beta} + L(\phi,t,x)^A = 0, 
\end{equation}

\textit{iii}) an identity derived from the above two systems which we shall call the subsidiary
evolution equation. It is a subset of the integrability conditions, and we assume has the form,

\begin{equation}
\label{eq:subsidiary}
\partial_t C^A = S(\phi,t,x)^{iA}{}_B\partial_i C^B + R(\phi,\partial \phi,t,x)^A{}_B C^B, 
\end{equation}

\end{definition}

Condition \textit{iii)} ensures that, if $C^A$ vanishes at the initial surface, 
hen its derivative off the initial surface also vanish. 
This is clearly a necessary condition for the constraints to be satisfied
away from the initial surface --it is part of the integrability conditions--
but in general it is not sufficient. If the integrability conditions
exist for a first order quasi-linear systems, then a subset of them must have the above form.
In particular, if a further condition is assumed (to be discussed below)  
it must be a linear equation on the constraint quantities, but that does
not play a role in the present discussion.
The above identity should arise by taking a time derivative of the quantities $C^A$ and using
the evolution equation for $\phi^{\alpha}$. Then re-expressing all quantities in terms of $C^A$.
If that is so we then say that the system is constrained and $C^A$ are the constraint quantities.
Failure of the identity implies the existence of further constraints.

Equation (\ref{eq:subsidiary}) is in general not unique, and so $S(\phi,t,x)^{Aa}{}_B$.
Thus, in this generality, nothing can be said and there are simple examples which show
that some subsidiary systems do not inherit any good property form the evolution systems
from which they arise. This happens if there is some differential identity between the constraint
quantities. That is, if there exists $W^n{}_A$ such that 

\begin{equation}
\label{eq:constraint_identity}
W^n{}_A\nabla_n C^A = F(C^B).
\end{equation}
Since Eq. (\ref{eq:constraint_identity}) is an identity for arbitrary values of 
$\phi^{\alpha}$ we must have that

\begin{equation} 
\label{eq:kernel}
W^{(n}{}_A K^{Am)}{}_{\alpha} = 0.
\end{equation}

The following example \cite{CS} illustrates the problem:


\begin{example}
Consider the evolution equation,

\begin{equation}
\partial_t \phi^{\alpha} = l^i \partial_i \phi^{\alpha}
\end{equation}
in flat space-time where $l^i$ is any non-vanishing vector. 
Let the constraints be:

\begin{equation}
C_i{}^{\alpha} = \partial_i \phi^{\alpha}.
\end{equation}
Then $K_i{}^{\alpha n}{}_{\beta} = \delta^{\alpha}{}_{\beta}\delta^n{}_i$.
Since partial derivatives conmute we have

\begin{equation}
\partial_{[j}C_{i]} = \partial{[j}\partial_{i]} \phi^{\alpha} = 0
\end{equation}
Thus, any field satisfying $W^{an}{}_{\beta} = W^{[an]}{}_{\beta}$. 
Gives rise to an identity, and in fact all of them satisfy 
Eq. (\ref{eq:kernel}), 
$W^{(n}{}_A K^{Am)}{}_{\alpha} = W^{(n|j|}_{\beta}K_j{}^{\beta m)}{}_{\alpha} =
W^{(nm)}_{\alpha} = 0$.
Thus, the set of subsidiary conditions is given by:

\begin{equation}
\partial_t C_i{}^{\alpha}{} = l^j \partial_j C_i{}^{\alpha} +  W_i^{\alpha[jn]}{}_{\beta} \partial_n C_j{}^{\beta}
\end{equation}
where $W_i^{\alpha[jn]}{}_{\beta}$ is arbitrary. 
Taking, for instance, $W_i^{\alpha[jn]}{}_{\beta} = 0$ we get a 
symmetric hyperbolic system, but taking instead, 
$W_i^{\alpha[jn]}{}_{\beta} = -2 \delta^{\alpha}{}_{\beta} \delta_i{}^{[j} l^{n]}$ we get a 
system which is only weakly hyperbolic.
\end{example}


To exclude this indeterminacy we require the following condition:

\vspace{5mm}
\textbf{Assumption:}  \textit{For any $\omega_a$, and any $\phi^{\alpha}$, $K(\phi,x)^{iA}{}_{\beta}\omega_i$ is
surjective.}
\vspace{5mm}

This ensures there is no nontrivial $W^{n}{}_A$ satisfying (\ref{eq:kernel}).
This condition ensures that all constraints are differential, 
as opposite to merely algebraic ones, and there are no superfluous ones that can be 
deduced from any differential identity.
This property also ensures that the subsidiary system is linear in the constraint quantities.

 Algebraic constraints are dealt in a different way  beforehand 
(basically by restricting the set of variables to those satisfying them), so in that respect this
is not a real loss of generality. 
But there are many important constrained systems which do not satisfy this condition, 
although in many cases one can choose a subset of constraints
which do satisfy the condition, and the rest has a trivial evolution. 
That is the case in the two examples (representing  Maxwell's, and Einstein's systems respectively) 
given below.
There are cases in which there are sets of measure zero for which the surjectivity property
fails, [In the example above instead of taking $C_i{}^{\alpha}$ as the set of constraints, just take 
the subset $C^{\alpha}:= s^i C_i{}^{\alpha}$, for
arbitrary constant vector $s^i$.], 
those have to be treated in grater detail, for it could happens that for those 
values one looses the properties one is looking for.


\begin{example}
\label{eg:Maxwell_2}  
Maxwell's equations (\ref{eqn:Maxwell}) as introduced in the first section,
for the electric field and vector potential. Two terms proportional to the constraints have
been added. They are parameterized respectively by the numbers $\alpha$ and $\beta$.
In order for this system to represent only Maxwell's equations constraints must be imposed, 
indeed the full constraints needed to make the system equivalent to Maxwell's equations
are given by 
$C_{ijk} = 2\partial_{[i}W_{j]k}=0$, $C=\partial^iE_i=0$. 
This set of constraints does not satisfy the surjectivity condition, but 
there is a subset of them, given by
$(C:=\partial^iE_i, C_i:= \partial^j W_{ij} - \partial_i W_k{}^k)$, whose $K^{An}_{\alpha}$
satisfies the surjectivity condition and furthermore has a closed integrability condition. 

\begin{eqnarray}
\partial_t C = -\alpha \partial^i C_i \nonumber \\
\partial_t C_i = -\beta \partial_i C.
\end{eqnarray}

Once this subset is shown to vanish, we can use the subsidiary equation
$\partial_t C_{ijk} = \beta e_{k[i} \partial_{j]}C$ to ensure the vanishing of the whole set of constraint equations.

\end{example}


\begin{example}
\label{eg:KTS}
The following symmetric hyperbolic system describes linearized general relativity 
in flat space-time~\footnote{
The analysis for the non-linear system is identical, since the frozen coefficient principal 
symbol is the same (up to the addition of trivial shift contribution and multiplication by the lapse).}
, \cite{EC}, \cite{KST}, \cite{CPSTR}

\begin{eqnarray}
\partial_t k_{ij} &=& - e^{kl}\partial_k f_{lij} \nonumber \\
\partial_t f_{kij} &=& - \partial_k k_{ij}
\end{eqnarray}
where $k_{ij} = k_{(ij)}$, and $f_{kij} = f_{k(ij)}$, and as before $e^{ij}$ is the the flat Euclidean metric. 
It admits a one parameter family of constraints: 
$(C_i := \partial^j(k_{ij} - e_{ij}k_k{}^k), C_{ij} := 
\partial^k f_{ijk} - \partial_j f_{il}{}^l + \alpha(\partial_j(f_{il}{}^l - f_l{}^l{}_i) 
- e_{ij}(\partial^n(f_{nl}{}^l - f_l{}^l{}_n))$.
For $\alpha \neq \frac{-1}{3}$ the corresponding linear map $K^{An}{}_{\alpha}$ is surjective, 
and the subsidiary system is given by:

\begin{eqnarray}
\partial_t C_i &=& -\partial^kC_{ki} \nonumber \\
\partial_t C_{ij} &=& - \partial_i C_j + \alpha(\partial_j C_i - e_{ij}\partial^kC_k)
\end{eqnarray}

This system has more constraints, but they are preserved (satisfy an homogeneous ordinary 
evolution equation), if the above ones vanish (see \cite{CPSTR}).

\end{example}


We now focus on the main result of this section.
Taking a time derivative of (\ref{eq:evolution}) and using (\ref{eq:constraints}), 
it is a short calculation to show that for (\ref{eq:subsidiary}) to be 
valid for arbitrary $\phi^{\alpha}$ the following algebraic identity must hold:

\begin{equation}
\label{eq:algebraic}
K^{(i|A|}{}_{\alpha} A^{j)\alpha}{}_{\beta} - S^{(i|A|}{}_B K^{j)B}{}_{\beta} = 0
\end{equation}

From this identity we can deduce the following results:

\begin{lemma}
Given any fixed non-vanishing co-vector $\omega_a$.
If $(\sigma, \phi^{\alpha})$ is an eigenvalue-eigenvector pair of $A^{i\alpha }{}_{\beta}\omega_i$ 
then 
$(\sigma, v^A = K^{iA}{}_{\alpha} \omega_i u^{\alpha})$, if $v^A$ is non-vanishing, is an  
eigenvalue-eigenvector pair of $ S^{iA}{}_B\omega_i$.
\end{lemma}

\begin{proof} 
Contract the above identity twice with $\omega_i$ and then with $u^{\alpha}$. 
\end{proof}

\begin{proposition} If the above \textbf{assumption} holds, then
if $A^{i \alpha}{}_{\beta}\omega_i$ has a complete set of eigenvectors so does 
have 
$S^{iA}{}_B\omega_i$. Furthermore the set of eigenvalues of $S^{iA}{}_B\omega_i$ is a 
subset of the set of eigenvalues of $A^{i\alpha}{}_{\beta}\omega_i$.
\end{proposition}

\begin{proof} If the eigenvalues of $A^{i\alpha}{}_{\beta}\omega_i$, $\{u^{\alpha}\}$ expand 
the whole space, then the set 
$\{v^A = K^{iA}{}_{\alpha} \omega_i u^{\alpha}\}$ expands the rank of 
$K^{iA}{}_{\alpha}\omega_i$, but this map is
assumed to be surjective, so they expand the whole space. 
Since they are all possible eigenvectors, and since they have
the same eigenvalues as the the set $\{u^{\alpha}\}$ from which they are built on, 
we conclude that all the eigenvalues of $S^{iA}{}_B\omega_i$ are also eigenvalues of 
$A^{i\alpha }{}_{\beta}\omega_i$. 
\end{proof}

In particular we have the following trivial consequences:

\begin{corollary} 
If $A^{i \alpha}{}_{\beta}\omega_i$ has a complete set of eigenvectors and 
if all their eigenvalues are real,  then so are the eigenvalues of $ S^{iA}{}_B\omega_i$.
\end{corollary}

\begin{corollary}
The characteristics of the subsidiary system are a subset of the characteristics of the evolution
system.
\end{corollary}

The above Proposition and its first Corollary imply the main result:

\begin{proposition} 
If the evolution system is strongly hyperbolic, then so is the subsidiary 
system.
\end{proposition}

Remarks:

\begin{itemize}
\item By strongly hyperbolic here we refer to the algebraic properties, we have not explored the
possibility that with these conditions one could conclude that if the symmetrizer of the evolution
system is smooth then the symmetrizer of the subsidiary system would also be smooth~\footnote{We refer to
the first section for further comments on this point.}.
That might very well be the case.

\item There remains the question as to whether a symmetric hyperbolic system would yield a
symmetric hyperbolic subsidiary system or the converse, namely if a non-symmetric but strongly
hyperbolic system could give rise to a symmetric hyperbolic subsidiary system.

In example \ref{eg:KTS} the system is strongly hyperbolic for all values of the parameter $\alpha$. 
But there are no symmetrizers when $\alpha < \frac{-1}{2}$ or $\alpha > 1$ for the subsidiary system. 
Thus a symmetric hyperbolic system gives a subsidiary system which is only strongly hyperbolic.

In example \ref{eg:Maxwell_2} the two parameter family system is strongly hyperbolic for all values of the parameters such that 
 $\alpha \beta > 0$. But no symmetrizer exists for $\alpha > 0$ or $\beta > \frac{-2}{3}$. 
Thus they are not symmetric hyperbolic.
Nevertheless, the constraint quantities, have a symmetric hyperbolic subsidiary system for 
all values of the parameters for which the above one is strongly hyperbolic.

\end{itemize}

\section{Acknowledgments}

I am thankfull from discussions on some of these topics with R. Geroch, G. Nagy, O. Ortiz, and
M. Tiglio. I thank the Caltech Visitors Program for the Numerical Simulation of
Gravitational Wave Sources, the Klavi Institute for Theoretical Physics Visitor Program: 
Gravitational Interaction of Compact Objects, and the Isaac Newton Institute for 
Mathematical Sciences, Cambridge Visitor Program: Hyperbolic Models in Astrophysics and Cosmology 
for their hospitality, where a portions of this
work was completed.

\end{document}